\documentclass[aps,amssymb,prb,twocolumn,showpacs]{revtex4-1}
\usepackage{graphicx}
\usepackage{epsfig,amsmath}
\begin{document}

\title{Single-electron heat diode}
\author{Tomi Ruokola$^1$}
\author{Teemu Ojanen$^2$}
\affiliation{$^1$ Department of Applied
Physics,  Aalto University, P.~O.~Box 11100,
FI-00076 Aalto, Finland}
\affiliation{$^2$ Low Temperature Laboratory, Aalto University, P.~O.~Box 15100,
FI-00076 Aalto, Finland }
\date{\today}
\begin{abstract}
We introduce a new functional nanoscale device, a single-electron heat
diode, consisting of two quantum dots or metallic islands coupled to
electronic reservoirs by tunnel contacts.  Electron transport through
the system is forbidden but the capacitive coupling between the two
dots allows electronic fluctuations to transmit heat between the
reservoirs. When the reservoir temperatures are biased in the forward
direction, heat flow is enabled by a four-step sequential tunneling
cycle, while in the reverse-biased configuration this process is
suppressed due to Coulomb blockade effects. In an optimal setup the
leakage heat current in the reverse direction is only a few percent of
the forward current.
\end{abstract}
 \bigskip
\maketitle

Understanding the thermal transport properties of nanoscale systems
offers insight into fundamental physics as well as opens up
possibilities for new applications.  In a solid-state environment heat
is generally carried by phonons and electrons, the latter offering
much more flexibility in manipulation and measurement. Therefore in
recent years there has been a great deal of interest to study heat
control with mesoscopic electronics, including devices for cooling,
\cite{rmp} thermoelectric power generation,\cite{linke,sanchez} and
thermal rectification.\cite{rect2}  These advances suggest that
in the near future thermal currents could be manipulated with a
similar level of versatility as electrons today. This
entails the construction of complicated heattronic circuits out of
elementary components.  One basic building block in electronics is a
diode, and its heattronic counterpart, a heat diode, is a two-terminal
device that allows energy to flow only in one direction. More precisely,
with a temperature bias in the forward direction, a heat current $J_+$
flows between the terminals, while reversing the temperatures makes
the current drop to $J_-$. For a proper diode operation these two
values should be separated by at least an order of magnitude, that is,
the rectification ratio $J_+/J_-$ should be at least of the order of
10. In this Rapid Communication we introduce a new heat diode design where this
requirement is easily fulfilled. 
That should be compared to
recent experimental demonstrations of asymmetric heat flow
\cite{rect1,rect2,rect3} where the ratio has been much below 2.
We also emphasize that although there is a considerable body of
theoretical literature on thermal rectification (see, for example,
Ref.~\onlinecite{segal} and references therein), it is mostly concerned with
simple model systems with no concrete realizations available, while
the present proposal can be straightforwardly fabricated and operated
with current experimental technology.

\begin{figure}[t]
\centering
\includegraphics[width=\columnwidth,clip]{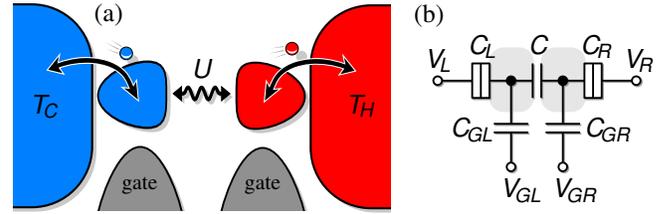}
\caption{(color online). (a) Schematic of the heat diode, shown here
with a forward bias.  Electrons
hop in and out of the dots and exchange heat through Coulomb
interaction $U$.  (b) Circuit diagram for the diode with capacitances
and external voltages indicated.  Shaded areas correspond to the dots.
}\label{fig1}
\end{figure}

\begin{figure}[t]
\centering
\includegraphics[width=\columnwidth,clip]{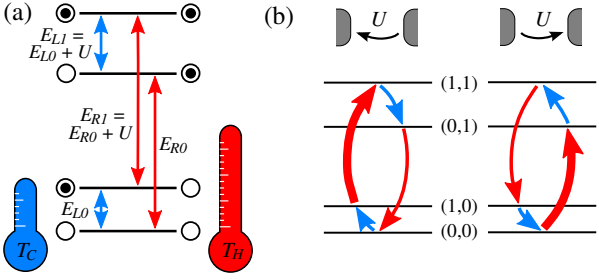}
\caption{(color online). (a) Energy-level diagram of the double-dot system.
(b) Sequential tunneling cycles transporting a heat quantum $U$ between
the reservoirs. The
processes denoted by the thick arrows become suppressed under reverse
bias. }\label{fig2}
\end{figure}

Our proposed device is a single-electron heat diode, consisting of two
semiconductor quantum dots or two metallic islands (collectively
called {\it dots}), coupled to different reservoirs ($L$ and $R$)
through tunnel barriers (Fig.~\ref{fig1}). The coupling between the
dots is purely capacitive so that there is no electron transport
through the system. However, with a temperature difference between the
reservoirs, thermal current is generated by electrons tunneling in and
out of the dots and exchanging energy through Coulomb interaction.
This energy transport mechanism has been previously studied in 
Ref.~\onlinecite{sanchez} in the context of a mesoscopic heat engine.
Here we extend the idea to produce a rectifying mechanism for
heat currents.
Let us fix the notation by
stipulating that in the forward direction the right reservoir is hot
and the left one is cold, with temperatures $T_H$ and $T_C$,
respectively.  The energy required for an electron to tunnel from
reservoir $\alpha$ into the adjoining dot is $E_{\alpha n}$, where $n$
is the occupation of the other dot. With two gate voltages the level
structure of the double dot can be tuned to such a regime where the
occupation of each dot may only be 0 or 1, and where the energies
$E_{Ln}$ are of the same order as $T_C$ and the energies $E_{Rn}$ are
much larger, at least as large as $T_H$ (Fig.~\ref{fig2}). Now one
quantum of heat, equal to the Coulomb interaction energy
$U=E_{L1}-E_{L0}=E_{R1}-E_{R0}$, can be transported from right to left
with the four-step cycle depicted on the left side of
Fig.~\ref{fig2}(b).
The time-reversed cycle carries heat in the other direction, and the
total current is the sum of these two contributions.  When the
temperatures are reversed, electrons from the right reservoir, now at
the low temperature $T_C$, are unable to tunnel into the adjoining dot
because the available thermal energy is much smaller than the required
energies $E_{Rn}$. Therefore the heat-carrying cycle is exponentially
suppressed and heat flow is blocked. This asymmetric Coulomb blockade
configuration is the origin of the diode effect.  The general idea of
producing rectification by coupling the left and right reservoirs to
different transitions of the central system has been previously
discussed in Ref.~\onlinecite{ojanen}.

The rest of the paper is
organized as follows.
First we present the device model in detail and
calculate the heat current in the
sequential tunneling
approximation. Then we show quantitatively how the diode effect arises in
the proposed device, considering two realizations, metallic islands
with a continuous spectrum and quantum dots with discrete states, and
conclude that their behavior is essentially identical. We analyze the conditions for optimal
diode operation and show how to obtain the required level structure
with a double-dot setup. Finally we consider the
experimental demonstration of the diode effect.

The system is modeled with
\begin{equation}
H = \sum_{\alpha=L,R}(H_\alpha + H_{D\alpha}+ H_{T\alpha}) + H_C,
\end{equation}
where $H_\alpha=\sum_{i\in\alpha}\varepsilon_i c_i^\dagger c_i$ is the
Hamiltonian for the reservoirs, $H_{D\alpha}=\sum_{i\in
  D\alpha}\varepsilon_{Di} c_{Di}^\dagger c_{Di}$ for the dots, and
$H_{T\alpha}=\sum_{i\in\alpha,j\in D\alpha} t_{i} c_{Dj}^\dagger c_{i}
+ {\rm h.c.}$ for tunneling between reservoirs and dots. $H_C$
contains the Coulombic charging and interaction energies.  We study a
parameter regime where the charge states of the double dot can be
truncated to the four lowest levels which we call $(0,0)$, $(1,0)$,
$(0,1)$, and $(1,1)$. The ground state is labeled $(0,0)$ and
$(n_L,n_R)$ is the state with $n_\alpha$ excitations in dot $\alpha$.
The particles occupying the dots in the excited states can be either
electrons or holes, and the Coulomb energy $U$ is positive or negative
depending on whether the two excitations in the $(1,1)$ state have
equal or opposite charges.  S\'anchez and B\"uttiker\cite{sanchez} use
a similar energy level structure to produce a heat engine. However,
their device requires energy filtering with single-level quantum dots,
whereas our diode principle does not depend on the dot density of
states and is therefore applicable also to metallic islands.

In the sequential tunneling approximation
heat current is calculated in terms of the rates $\Gamma_{\alpha
n}^{(\sigma)}$ for tunneling between reservoir $\alpha$ and the
adjoining dot, with $n$ being the occupation of the other dot and
$\sigma=1$ for tunneling into the dot, $\sigma=0$ for tunneling into
the reservoir. The probabilities $P_{n_Ln_R}$ to be in state $(n_L,n_R)$
are obtained from a steady-state master equation, and the heat current through the
system, equal to the net energy extracted from the right reservoir, is
then given by\cite{sanchez}
\begin{eqnarray}\label{jseq}
J&=&E_{R0}(P_{00}\Gamma_{R0}^{(1)} - P_{01}\Gamma_{R0}^{(0)})+
E_{R1}(P_{10}\Gamma_{R1}^{(1)} - P_{11}\Gamma_{R1}^{(0)}) \nonumber\\
&=&U\tilde{\Gamma}^{-3}
(\Gamma_{L0}^{(1)}\Gamma_{R1}^{(1)}\Gamma_{L1}^{(0)}\Gamma_{R0}^{(0)}-
\Gamma_{R0}^{(1)}\Gamma_{L1}^{(1)}\Gamma_{R1}^{(0)}\Gamma_{L0}^{(0)}),
\end{eqnarray}
where the two terms on the second line correspond to the two cycles in
Fig.~\ref{fig2}(b), the normalization is $\tilde{\Gamma}^{3} =
\sum_{\alpha=L,R}\sum_{n,\sigma=0,1}
\Gamma_{\bar{\alpha}n}^{(\sigma)}\Gamma_{\alpha\bar{\sigma}}^{(n)}
(\Gamma_{\alpha\sigma}^{(\sigma)} +
\Gamma_{\alpha\sigma}^{(\bar{\sigma})})$, and a top bar denotes the
other possible value, for example, $\bar{0}=1$, $\bar{L}=R$.  Fermi
golden rule gives the tunneling rates as $\Gamma_{\alpha n}^{(\sigma)}
= 2\pi\Gamma_\alpha F_\alpha\big((-1)^{\bar{\sigma}}E_{\alpha
n}\big)$, where
$\Gamma_\alpha=\sum_{i\in\alpha}|t_i|^2\delta(\varepsilon_i)$ is the
tunneling strength for junction $\alpha$, assumed to be energy
independent. For metallic islands with a continuous spectrum
the effective reservoir occupation is
$F_\alpha(E) = E\nu_{D\alpha}n_\alpha(E)$, where
$\nu_{D\alpha}=\sum_{i\in D\alpha}\delta(\varepsilon_{Di})$
is the island density of states and $n_\alpha$ is the Bose function.
For quantum dots with a single discrete level $\varepsilon_{D\alpha}$
we have $F_\alpha(E) = f_\alpha(E)$, the Fermi function for reservoir
$\alpha$.
Substituting these in
Eq.~(\ref{jseq}) gives for a symmetric structure
($\Gamma_L=\Gamma_R\equiv\Gamma$)
\begin{equation}\label{j}
J = A^{-1}U\Gamma[n_R(U)-n_L(U)].
\end{equation}
The diode effect is contained in the asymmetric weight
function $A$, given for quantum dots as
\begin{equation}\label{a1}
A =\frac{ 2[1+n_L(U)][1+n_R(U)] }{
f_{L0}(1-f_{L1})f_{R0}(1-f_{R1}) } -2
\end{equation}
and for metallic islands as
\begin{equation}\label{a2}
A = \nu_D^{-1}\sum_{\substack{\alpha=L,R\\n,\sigma=0,1}}
\frac{[\sigma+n_\alpha(U)][\bar{\sigma}+n_{\bar{\alpha}}(U)]}{
E_{\alpha n}[\delta_{n\sigma}+n_{\alpha n}]
[\bar{n}-(-1)^n f_{\bar{\alpha}\delta_{n\sigma}}]},
\end{equation}
where $f_{\alpha n}=f_\alpha(E_{\alpha n})$,
$n_{\alpha n}=n_\alpha(E_{\alpha n})$, and $\nu_D=\nu_{DL}=\nu_{DR}$.

The performance of the diode can be assessed by analyzing two key
quantities, the heat current under forward bias, $J_+$, and the ratio
of forward and reverse currents, $J_+/J_-$. Forward bias is defined as
the right reservoir having the high temperature $T_H$ and the left
reservoir having the low temperature $T_C$; under reverse bias the
temperatures are reversed.  We start the analysis by making
simplifications to Eqs.~(\ref{j}) and (\ref{a1}).  First note that the
constant term in Eq.~(\ref{a1}) is practically always small compared
to the total $A$ and can be neglected. The four Fermi factors in the
denominator of $A$ are then divided into two groups: those with a
value larger than $\frac 1 2$ are taken to be constants while the rest
are approximated by exponentials.  The device has two qualitatively
different operating regimes, one with a positive and the other with a
negative $U$, and we first consider $U>0$. In this case we have
$1-f_{\alpha 1}\approx 1$, giving the dependence of forward current on
the energies $E_{\alpha 0}$ as $J_+ \propto
e^{-\frac{E_{L0}}{T_C}-\frac{E_{R0}}{T_H} }$, and the rectification
ratio as
\begin{equation}\label{posu}
J_+/J_-\approx e^{(T_C^{-1}-T_H^{-1})(E_{R0} -E_{L0})}.
\end{equation}
We see that $(E_{R0} -E_{L0})/T_C$ should be large for efficient
rectification while $E_{L0}/T_C$ and $E_{R0}/T_H$ should not be much
above unity for a large forward current.  Also, the difference in the
temperatures should preferably be in the range
$T_H/T_C\gtrsim2$.  Since by definition $(0,0)$ is the ground state,
both energies $E_{\alpha 0}$ must be positive, and thus the optimal
value of $E_{L0}$ is zero. The energy $E_{R0}$ can then be chosen to
achieve the desired balance between forward current and diode
efficiency.  The interaction energy $U$ has only a small effect on
rectification but the forward current depends on it strongly. When
$U\gg T_H$, the Bose functions in Eq.~(\ref{j}) suppress the current
exponentially, while in the limit $U\ll T_H$ the Bose functions in
Eq.~(\ref{a1}) give a similar suppression. Thus the optimal value of
$U$ is of the order of $T_H$.

When considering the other operating regime, $U<0$, we can restrict
attention to the interval $E_{L0}<|U|<E_{R0}$ since
a large $|U|$ would push the $(1,1)$ state
below the ground state, while the
above approximations for positive $U$ are actually valid for $U>-E_{L0}$
and therefore the
(uninteresting) case of $|U|<E_{L0}$ is contained in the
previous discussion. Now we can approximate $1-f_{R1}\approx 1$
while the three other Fermi factors in $A$ are taken as exponentials.
Then the current depends on the denominator of $A$ as $J_+ \propto
e^{-\frac{|U|}{T_C}-\frac{E_{R0}}{T_H} }$, and the rectification ratio
is
\begin{equation}\label{negu}
J_+/J_-\approx e^{(T_C^{-1}-T_H^{-1})(E_{R0} -|U|)}.
\end{equation}
Here $|U|$ has replaced $E_{L0}$ in both expressions and therefore
the device operation is almost independent of $E_{L0}$ in this regime.
Numerical calculations show that the forward current is maximized
when $E_{L0}\approx 2\,T_C$.

\begin{figure}[t]
\centering
\includegraphics[width=\columnwidth,clip]{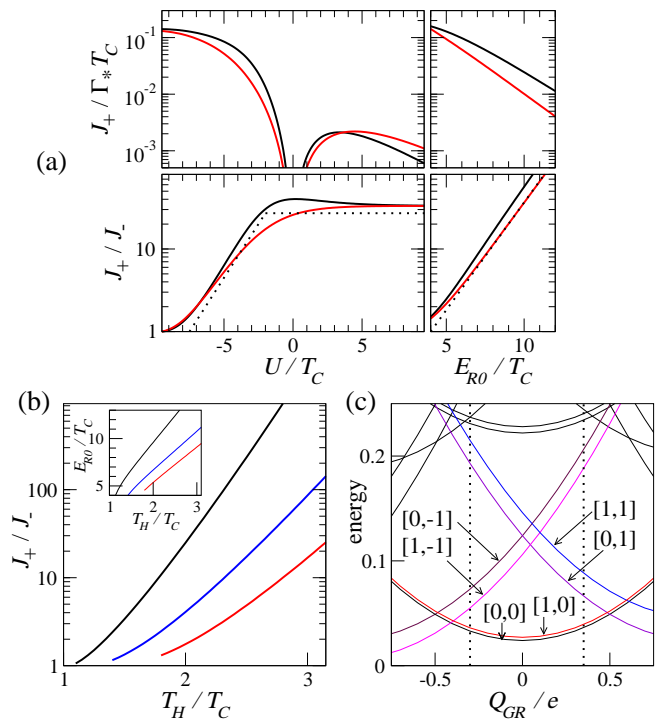}
\caption{(color online). (a) Forward current $J_+$ and the rectification ratio
$J_+/J_-$ as a function of $U$ and $E_{R0}$. Black curves correspond to quantum dots
($\Gamma_*=\Gamma$), red for metal islands ($\Gamma_*=\Gamma\nu_D
E_{R0}$).  Dotted lines are the estimates from Eqs.~(\ref{posu}) and
(\ref{negu}). The parameters are $T_H=2.5\,T_C$, $E_{L0}=2\,T_C$,
$E_{R0}=7.5\,T_C$ (left panels) and $U=-4\,T_C$ (right panels).
(b) Maximum rectification as a function of thermal
bias $T_H/T_C$ for fixed values of forward current.
The three curves correspond to, from top to bottom,
$J_+/\Gamma_*T_C=0.01$ (black), 0.05 (blue), 0.1 (red). For each data
point the parameters $E_{L0}$, $E_{R0}$ and $U$ are optimized to give
maximum rectification for the given current. Optimal values of
$E_{R0}$ are shown in the inset, for $E_{L0}/T_C$ they are in the
range $2\dots3$, for $U/T_C$ in the range $-5.5\dots-3.5$.  (c)
Charging energies for the double dot as a function of the right gate
charge $Q_{GR}$.  The curves correspond to different charge states,
with the relevant ones marked as $[N_L,N_R]$, where $N_\alpha$
is the number of electrons in dot $\alpha$. Dotted lines highlight two
values for $Q_{GR}$ where the heat diode could be successfully
operated, that is, the left and right transitions are well separated,
and the lowest four levels are well separated from higher levels.
Here the left gate charge is $Q_{GL} = .48\,e$, the capacitance
constants are $g_L=g_R=.1$, and the unit of energy is
$e^2/(2C(1-g_Lg_R))$.  }\label{fig3}
\end{figure}

Above we considered only the quantum dot setup, characterized by
Eq.~(\ref{a1}). However, the same approximations and conclusions can
be derived from Eq.~(\ref{a2}) by taking into account only the
dominant two terms of the sum, that is, those terms where both
occupation factors in the denominator are close to zero.
The conclusions can also be verified by a numerical analysis of
the full equations, as is done in Fig.~\ref{fig3}.
We see that both positive and negative $U$ can produce
similar levels of rectification but the forward current is
larger for $U<0$ and therefore we concentrate on that regime.
To have an efficient diode with $J_+/J_-> 10$,
we see that
unless a temperature
bias of $T_H/T_C>3$ is available, the forward current cannot be
larger than about $10^{-1}\,\Gamma_*T_C$, where $\Gamma_*=\Gamma$ for
quantum dots and $\Gamma_*=\Gamma\nu_DE_{R0}$ for metal islands.
On the other hand, if a current level of $10^{-2}\,\Gamma_*T_C$
is sufficient, considerable rectification takes place already
for $T_H/T_C<2$.

The validity of the above analysis requires that the dots are not
coupled too strongly to the leads. Since we require that only
the lowest four levels are accessible, the tunneling rates should be
smaller than the energies of the neglected higher-lying states.  The
energy scale for the forbidden transitions is set by $E_{R0}$ and thus
we must have $2\pi\Gamma_*\ll E_{R0}$. This coincides with the
requirement that the junction conductances must be well below the
conductance quantum.  On the other hand, coupling strength must also
be limited to prevent leakage currents due to cotunneling effects.  As
already discussed, the suppression of current under reverse bias is
due to the fact that the low-energy particles are unable to tunnel
into the right dot. This blockade can be lifted by coherent
two-electron processes where particles tunnel through both junctions
simultaneously and most of the required energy is contributed by the
left reservoir. There are two such cotunneling processes: if the
system is initially in the $(0,0)$ state, particles tunnel into both
dots and the final state is $(1,1)$; the other possible process leads
from $(1,0)$ to $(0,1)$. In the supplementary material\cite{suppl} we
present a detailed calculation of these processes and arrive at the
conclusion that they can be neglected when $\Gamma_*$ is bounded
as above.

Let us now consider how to actually realize the required energy level
structure in a double-dot setup.  For a metallic island, the tunneling
energy $E_{\alpha n}$ is simply the difference of electrostatic
charging energies of the initial and final states, for the quantum dot
system one must also add the bare energy of the discrete
single-particle level, $\varepsilon_{D\alpha}$.  Using the notation of
Fig. 1(b), the charging energy for dots with charges $Q_L$ and $Q_R$
is
\begin{equation}\label{ech}
\begin{split}
&E_{ch}(Q_L,Q_R)=\\
&\frac{g_L(Q_L+Q_{GL})^2+g_R(Q_R+Q_{GR})^2+
2g_Lg_RQ_LQ_R}{2C(1-g_Lg_R)}
\end{split}
\end{equation}
with interaction constants $g_\alpha=C/(C+C_\alpha+G_{G\alpha})$ and gate
charges $Q_{G\alpha} = C_{G\alpha}(V_{G\alpha}-V_\alpha) +
g_{\bar{\alpha}}C_{\bar{\alpha}}(V_{\bar{\alpha}}-
V_\alpha)+g_{\bar{\alpha}}C_{G\bar{\alpha}}(V_{G\bar{\alpha}}-V_\alpha)$.
We see that the level
structure is controlled by two gate charges which in turn depend on three
voltages, and therefore in a practical implementation one of the voltages
can be discarded.
If in the ground state $(0,0)$ the dots have charges $Q_{L0}$ and
$Q_{R0}$, then in the excited dots the charges are $Q_{L0}+q_L$
and $Q_{R0}+q_R$, where the $q_\alpha$ are $-e$ or $+e$
depending on whether the added excitation is an electron or a hole.
From Eq.~(\ref{ech}) we then see that the magnitude of $U$ is determined
by the capacitances as $|U|=e^2/[C(g_L^{-1}g_R^{-1}-1)]$ while the
energies $E_{L0}$ and $E_{R0}$ as well as the sign of $U$ can be
chosen with the gate charges. The energies $E_{\alpha 0}$ are
\begin{eqnarray*}
E_{L0} &=& E_{ch}(Q_{L0}+q_L, Q_{R0}) - E_{ch}(Q_{L0}, Q_{R0}) -
\frac{q_L}{e}\varepsilon_{DL}\\
E_{R0} &=& E_{ch}(Q_{L0}, Q_{R0}+q_R) - E_{ch}(Q_{L0}, Q_{R0}) -
\frac{q_R}{e}\varepsilon_{DR}
\end{eqnarray*}
where the single-particle levels $\varepsilon_{D\alpha}$ vanish for
metallic islands.  If the truncation of the system to the four lowest
charge states is to be valid, all higher states must have an energy
much above all $E_{\alpha n}$. Of course the levels $E_{\alpha n}$
must themselves have the structure discussed above.  Figure
\ref{fig3}(c) shows two concrete examples of how to obtain appropriate
level diagrams based on the above considerations.

Experimental demonstration of the diode effect must take place at
around 1 K or below. First, the requirement of large absolute bias,
$T_H/T_C\sim2$, can only be attained at low temperatures.  Second, since
the device is based on Coulomb blockade physics, the charging energies
attainable in mesoscopic structures set the maximum operating
temperature to about 1 K.  Third, in order to observe the thermal
current through the diode, heat transport by phonons must be
suppressed, and this generally takes place at sub-Kelvin temperatures.
\cite{meschke}  In a realistic setup we could have $T_C=100$~mK,
$T_H=250$~mK, and $\Gamma_*=0.5\,T_C$, giving
$\Gamma_*T_C\approx10$~fW. Referring to Fig.~\ref{fig3}, we see that it is
then possible to have a forward current of the order of 1~fW with the
reverse current being a few percent of this value. The forward current is
large enough to be measured with state-of-the-art thermometry.
\cite{peltonen}

In summary, we propose a new device, a single-electron heat diode,
which can be realized by a double-dot system connecting two electronic
reservoirs at different temperatures. The device and required
operation scheme are routinely realized with currently existing
technology. We explored the rectification performance of the device in
detail and showed that with experimentally measurable current
levels it is possible to have a rectification ratio well above 10,
even up to about 100,
making the present device the first concrete proposal for an
efficient heat diode.

The authors would like thank Antti-Pekka Jauho, Jukka Pekola and
Joonas Peltonen for useful comments. One of the authors (T.O.)
acknowledges the Academy of Finland for financial support.

\setcounter{equation}{0}
\onecolumngrid
\section*{Supplementary material: Virtual two-particle tunneling rates for
the single-electron heat diode}

As explained in the main text, heat flow under reverse thermal bias is
blocked because the golden-rule rates for hopping into the right dot
are exponentially suppressed. This blockade can be lifted by coherent
tunneling of electrons through both junctions simultaneously, with
most of the energy contributed by the hot left reservoir.  There are
two different processes which can lift the blockade: if the system is
initially in the $(0,0)$ state, particles can tunnel into both dots
and the system ends up in the $(1,1)$ state. If, on the other hand, we
start with the $(1,0)$ state, one particle hops out of the left dot
and another one into the right dot, leaving the system in the $(0,1)$
state.  Here we calculate the rates for these virtual transitions and
show that in the parameter range relevant for diode operation they are
negligible compared to sequential processes.  The rates for
two-particle transitions can be calculated from the general expression
\cite{bruus}
\begin{align}\label{formal}
\Gamma^{(2)}=2\pi\sum_{f_b,i_a}
|\langle f_b|H_{T}\frac{1}{E_{i_a}-H_0}H_{T}|i_a\rangle|^2W_{i_a}
\delta(E_{f_b}-E_{i_a}),
\end{align}
where $H_T = H_{TL}+H_{TR}$
is the tunneling Hamiltonian, $H_0=H-H_T$ the uncoupled Hamiltonian,
and $W_{i_a}$ the probability weight for the initial
states.
For the process $(0,0)\to(1,1)$ the final state
is related to the initial state by $|f_b\rangle=c^\dagger_{DL}c_L
c^\dagger_{DR}c_R|i_a\rangle$. Note that this only applies to the case
when the excitations for both dots are electrons. If, for instance,
the excitation tunneling into the right dot is a hole, one should make the
change $c^\dagger_{DR}c_R \to c_{DR}c^\dagger_R$. However, the formulas
below will be identical in all the different cases.
Writing the rate explicitly for metallic islands
yields
\begin{equation}\begin{split}\label{island}
\Gamma^{(2)}_{00\to11} =\ & 2\pi\Gamma_L\Gamma_R\nu_{DL}\nu_{DR}
\int d\varepsilon_L d\varepsilon_{DL} d\varepsilon_R d\varepsilon_{DR}\,
f_L(\varepsilon_{L})[1-f_L(\varepsilon_{DL})]f_R(\varepsilon_R)[1-f_R(\varepsilon_{DR})]\times\\
&\left|\frac{1}{\varepsilon_L-\varepsilon_{DL}-E_{L0} -\frac{i}{2}\Gamma_{10}} +
 \frac{1}{\varepsilon_R-\varepsilon_{DR}-E_{R0}-\frac{i}{2}\Gamma_{01}}
\right|^2
\delta(\varepsilon_L+\varepsilon_R-\varepsilon_{DL}-\varepsilon_{DR}-E_{L0}-E_{R0}-U).
\end{split}\end{equation}
The two energy denominators correspond to the two different orderings in which
the electron hoppings can occur. To regularize the divergences of the denominators
we have introduced the inverse lifetimes $\Gamma_{n_Ln_R}$ of
the state $(n_L,n_R)$, calculated with the lowest-order golden rule. Thus we have,
for example, $\Gamma_{10} = \Gamma_{L0}^{(0)} + \Gamma_{R1}^{(1)}$.
In the case of quantum dots the corresponding rate is
\begin{equation}\begin{split}\label{qdot}
\Gamma^{(2)}_{00\to11} = 2\pi\Gamma_L\Gamma_R
\int d\varepsilon_L d\varepsilon_R\,
f_L(\varepsilon_L)f_R(\varepsilon_R)
\left|\frac{1}{\varepsilon_L-E_{L0}-\frac{i}{2}\Gamma_{10}} +
\frac{1}{\varepsilon_R-E_{R0}-\frac{i}{2}\Gamma_{01}}
\right|^2
\delta(\varepsilon_L+\varepsilon_R-E_{L0}-E_{R0}-U).
\end{split}\end{equation}
For the other relevant cotunneling process, from $(1,0)$ to $(0,1)$, the
metallic system has
\begin{equation}\begin{split}\label{island2}
\Gamma^{(2)}_{10\to01} =\ & 2\pi\Gamma_L\Gamma_R\nu_{DL}\nu_{DR}
\int d\varepsilon_L d\varepsilon_{DL} d\varepsilon_R d\varepsilon_{DR}\,
[1-f_L(\varepsilon_{L})]f_L(\varepsilon_{DL})f_R(\varepsilon_R)[1-f_R(\varepsilon_{DR})]\times\\
&\left|\frac{1}{\varepsilon_{DL}-\varepsilon_L-E_{L0} -\frac{i}{2}\Gamma_{00}} +
 \frac{1}{\varepsilon_R-\varepsilon_{DR}-E_{R1}-\frac{i}{2}\Gamma_{11}}
\right|^2
\delta(-\varepsilon_L+\varepsilon_R+\varepsilon_{DL}-\varepsilon_{DR}+E_{L0}-E_{R0}),
\end{split}\end{equation}
and analogously for the quantum dot setup.

Equations (\ref{island}) and (\ref{island2}) suggest that for the metallic
system it is useful to define new integration variables $\varepsilon^\prime_\alpha=
\varepsilon_\alpha- \varepsilon_{D\alpha}$. Then with the help of the identities
$f_\alpha(\varepsilon^\prime_\alpha+\varepsilon_{D\alpha})[1-f_\alpha(\varepsilon_{D\alpha})]=
n_\alpha(\varepsilon_{D\alpha})[f_\alpha(\varepsilon_{D\alpha})-
f_\alpha(\varepsilon^\prime_\alpha+\varepsilon_{D\alpha})]$
and
$\int d\varepsilon_{D\alpha}\,[f_\alpha(\varepsilon_{D\alpha})-
f_\alpha(\varepsilon^\prime_\alpha+\varepsilon_{D\alpha})] = \varepsilon^\prime_\alpha$
the variables $\varepsilon_{D\alpha}$ can be integrated out.
For both metallic and quantum dot cases the delta function can be used to
 eliminate another of the remaining integration variables,
and then all the tunneling rates can be expressed with a single equation:
\begin{equation}\label{common}
\Gamma^{(2)}_{\Delta} = 2\pi\Gamma_L\Gamma_R U^2
\int d\varepsilon\,
F_L(\Delta-\varepsilon)F_R(\varepsilon)
\left|\frac{1}{\varepsilon-E_{R0}-\frac{i}{2}\Gamma_{0}}\right|^2
\left|\frac{1}{\varepsilon-E_{R1}-\frac{i}{2}\Gamma_{1}}\right|^2,
\end{equation}
where $F_\alpha(\varepsilon) = f_\alpha(\varepsilon)$ for quantum
dots and $F_\alpha(\varepsilon) = \varepsilon\nu_{D\alpha}n_\alpha(\varepsilon)$
for a metallic system.
For the $(0,0)\to(1,1)$ transition the parameters
are $\Delta = E_{L0}+E_{R0}+U$, $\Gamma_0=\Gamma_{01}$, and
$\Gamma_1=\Gamma_{10}$, and for the
$(1,0)\to(0,1)$ transition they are
$\Delta = E_{R0}-E_{L0}$, $\Gamma_0=\Gamma_{00}$, and
$\Gamma_1=\Gamma_{11}$.

It is important to notice that Eq.~(\ref{common})
gives the total transition rates, including both sequential tunneling and
cotunneling.\cite{nasse} Sequential contribution comes from the resonance peaks
at $\varepsilon=E_{R0}$ and $\varepsilon=E_{R0}$, while the rest, due to
the peak in the function $F_L(\Delta-\varepsilon)F_R(\varepsilon)$, is
the cotunneling contribution.
The sequential rate can therefore be calculated by taking the limit
$|\varepsilon-E-\frac{i}{2}\Gamma|^{-2}
\to \frac{2\pi}{\Gamma}\delta(\varepsilon-E)$.
We get
\begin{equation}
\Gamma^{(2)}_{00\to11,{\rm seq}} = \frac{\Gamma_{R0}^{(1)}\Gamma_{L1}^{(1)}}{\Gamma_{01}}
+\frac{\Gamma_{L0}^{(1)}\Gamma_{R1}^{(1)}}{\Gamma_{10}};\quad
\Gamma^{(2)}_{10\to01,{\rm seq}} = \frac{\Gamma_{R1}^{(1)}\Gamma_{L1}^{(0)}}{\Gamma_{11}}
+\frac{\Gamma_{L0}^{(0)}\Gamma_{R0}^{(1)}}{\Gamma_{00}}.
\end{equation}
This is exactly the result one would expect; for example, the first term
in the $(0,0)\to(1,1)$ process is the rate of the $(0,0)\to(0,1)$
transition, multiplied by the probability that the next transition leads to $(1,1)$.
The second term similarly gives the rate for the process
$(0,0)\to(1,0)\to(1,1)$.

To calculate the cotunneling rate we note that for $\exp(\Delta/T_H)\gg1$ we
can approximate $F_L(\Delta-\varepsilon)F_R(\varepsilon)\approx
e^{-\Delta/T_H}(e^{\varepsilon(T_C^{-1}-T_H^{-1})}+e^{-\varepsilon/T_H})^{-1}$
for the quantum dot system. This function has a peak at $|\varepsilon|\lesssim T_C$
and it decays exponentially with $1/(T_C^{-1}-T_H^{-1})$ and $T_H$ for positive
and negative $\varepsilon$, respectively. In this region we can approximate
$E_{Rn}-\varepsilon\approx E_{Rn}$ and therefore
the cotunneling contribution can be extracted by setting
$|\varepsilon-E-\frac{i}{2}\Gamma|^{-2}\to E^{-2}$
in Eq.~(\ref{common}). The result is
\begin{equation}
\Gamma^{(2)}_{\Delta,{\rm cot}}=
\frac{2\pi^2\Gamma_L\Gamma_R U^2 T_C}{
E_{R0}^2E_{R1}^2\sin\pi\frac{T_C}{T_H}}e^{-\Delta/T_H}
\end{equation}
for quantum dots. Similar considerations apply also in the metallic
case, and the cotunneling rate is
\begin{equation}
\Gamma^{(2)}_{\Delta,{\rm cot}}=
\frac{2\pi^3\Gamma_L\Gamma_R\nu_{DL}\nu_{DR} U^2 T_C^2\Delta}{
E_{R0}^2E_{R1}^2\sin^2\pi\frac{T_C}{T_H}}e^{-\Delta/T_H}.
\end{equation}
The relevance of cotunneling can now be estimated by comparing
the magnitudes of $\Gamma^{(2)}_{\rm cot}$ and
$\Gamma^{(2)}_{\rm seq}$. For a symmetric system,
$\Gamma_L=\Gamma_R\equiv\Gamma$, with $U\gtrsim T_H$,
we have 
\begin{equation}
\frac{\Gamma^{(2)}_{\rm cot}}{\Gamma^{(2)}_{\rm seq}}\approx
\frac{\pi\Gamma_{\rm eff} U^2 T_C}{
E_{R0}^2E_{R1}^2}e^{E_{L0}/T_H}e^{E_{R0}(T_C^{-1}-T_H^{-1})},
\end{equation}
where $\Gamma_{\rm eff}=\Gamma$ for quantum dots and
$\Gamma_{\rm eff}=\pi\Gamma\nu_D T_C$
for metal islands. We have used the rates for the $(1,0)\to(0,1)$
transition since the ratio would be smaller for the $(0,0)\to(1,1)$
transition. For $U\lesssim -T_H$ the latter process dominates and we have
\begin{equation}
\frac{\Gamma^{(2)}_{\rm cot}}{\Gamma^{(2)}_{\rm seq}}\approx
\frac{\pi\Gamma_{\rm eff} U^2 T_C}{
E_{R0}^2E_{R1}^2}e^{-E_{L0}/T_H}e^{E_{R1}(T_C^{-1}-T_H^{-1})}.
\end{equation}
For the parameter values corresponding to the plots in
Fig.~3 of the main text, with $\Gamma_{\rm eff}<T_C$,
these ratios are smaller than 10\%, and therefore
we conclude that cotunneling effects are negligible for
the proposed device.

\end{document}